\documentclass[12pt]{article}
\usepackage[varg]{txfonts}
\usepackage{cite} % collapse multiple cites, eg: 1,2,3,4,5 -> 1-5
\usepackage[scaled=0.9]{helvet}
\usepackage{graphicx}

\usepackage{url}  % typeset url's
\usepackage{array}

\evensidemargin\oddsidemargin
\parskip=\medskipamount

\makeatletter
\long\def\@makecaption#1#2{%
  \vskip\abovecaptionskip
  \sbox\@tempboxa{\small{\bfseries #1} \  #2}%
  \ifdim \wd\@tempboxa >\hsize
    \small{\bfseries #1} \  #2\par
  \else
    \global \@minipagefalse
    \hb@xt@\hsize{\hfil\box\@tempboxa\hfil}%
  \fi
  \vskip\belowcaptionskip}
\renewcommand\section{\@startsection {section}{1}{\z@}%
      {-3.25ex\@plus -1ex \@minus -.2ex}%
      {1ex \@plus .2ex}%
      {\normalfont\large\sffamily\bfseries}}
\renewcommand\subsection{\@startsection{subsection}{2}{\z@}%
      {-3ex\@plus -1ex \@minus -.2ex}%
      {0.5ex \@plus .2ex}%
      {\normalfont\normalsize\sffamily\bfseries}}
\renewcommand\subsubsection{\@startsection{subsubsection}{3}{\z@}%
      {-3ex\@plus -1ex \@minus -.2ex}%
      {0.25ex \@plus .2ex}%
      {\normalfont\normalsize\sffamily\bfseries}}
\renewcommand\paragraph{\@startsection{paragraph}{4}{\z@}%
      {3ex \@plus1ex \@minus.2ex}%
      {-1em}%
      {\normalfont\normalsize\sffamily\bfseries}}
\renewcommand\subparagraph{\@startsection{subparagraph}{5}{\z@}%
      {1ex \@plus.5ex \@minus .2ex}%
      {-1em}%
      {\normalfont\normalsize\sffamily\bfseries}}
\makeatother

\def\ibar{\overline I_0}
\def\qsqmax{q^2_\mathrm{max}}
\def\sth{s_\mathrm{th}}
\def\kl#1{K_{l#1}}
\renewcommand{\vec}[1]{\mathbf{#1}}
\def\mi{\mathrm{i}}
\def\e{\mathrm{e}}
\def\gev{\,\mathrm{GeV}}
\def\mev{\,\mathrm{MeV}}
\def\d{\mathrm{d}} % upright d for integration measure

\begin{document}
\begin{flushright}
SHEP--0628
\end{flushright}

\begin{center}\Large\bfseries\sffamily
Elastic $s$-wave $B\pi$, $D\pi$, $DK$ and $K\pi$ Scattering from
Lattice Calculations of Scalar Form Factors in Semileptonic Decays
\end{center}

\begin{center}
\textbf{\textsf{Jonathan M Flynn${}^\mathrm{a}$ and Juan
  Nieves${}^\mathrm{b}$}}\\[2ex]
${}^\mathrm{a}$School of Physics and Astronomy, University of
  Southampton\\
  Highfield, Southampton SO17~1BJ, UK\\
${}^\mathrm{b}$Departamento de F\'isica At\'omica, Molecular y
  Nuclear, Universidad de Granada,\\
  E--18071 Granada, Spain
\end{center}
\medskip

\begin{quote}
\begin{center}\textbf{\textsf{Abstract}}\end{center}
We show how theoretical, principally lattice, calculations of the
scalar form factors in semileptonic pseudoscalar-to-pseudoscalar
decays can be used to extract information about the corresponding
elastic $s$-wave scattering channels. We find values for the
scattering lengths $m_\pi a = 0.179(17)(14)$, $0.26(26)$ and $0.29(4)$
for elastic $s$-wave isospin-$1/2$ $K\pi$, $B\pi$ and $D\pi$ channels
respectively. We also determine phase shifts. For the $DK$ channel we
find hints that there is a bound state which can be identified with
the recently discovered $D_{s0}^+(2317)$.
\end{quote}

\section{Introduction}

The Omn\`es representation of form factors has been widely used in
descriptions of kaon
decays~\cite{Gasser:1984ux,Omnes_01_kaon,Jamin:2001zq,Ananthanarayan:2004xy}.
Usually, phase shift information or chiral perturbation theory $K\pi$
scattering amplitudes are taken as input to determine the scalar form
factor in $K_{l3}$ decays. Here we explore what can be inferred about
the $s$-wave isospin-$1/2$ $K\pi$ phase shift from theoretical lattice
determinations of this form factor below $\qsqmax = (m_K-m_\pi)^2$.
Direct lattice determinations of scattering observables face serious
difficulties~\cite{Maiani:1990ca,Ciuchini:1996mq}, but nonetheless
phase shift information can be extracted from finite-volume effects in
appropriate correlation functions~\cite{Lellouch:2000pv,Lin:2001ek}.
Results in the meson sector have appeared for $\pi\pi$ scattering in
the isospin-$2$ channel (see~\cite{Aoki:2005uf} and references
therein) and recently for $K\pi$ scattering in the isospin-$3/2$
channel~\cite{Beane:2006gj}\footnote{Reference~\cite{Beane:2006gj}
uses NLO chiral perturbation theory to extract the phase shift for the
isospin-$1/2$ channel in addition.}. Given this situation, we believe
it is worthwhile to explore the alternative route proposed above. We
also extend the discussion to encompass semileptonic decays of
heavy-light pseudoscalar mesons, $H=B,D$, to pions and kaons, and the
related elastic $s$-wave scattering reactions.

The form factor obtained from the Omn\`es representation becomes less
sensitive to the details of the phase shift above threshold as the
number of subtractions increases. This feature can be exploited in two
ways. By using a large enough number of subtractions, one can
determine the form factor without relying on any detailed knowledge of
the phase shift apart from its value at threshold.
Reference~\cite{Flynn:2006vr} used this to determine $|V_{ub}|$ from
$f_+$ in semileptonic $B\to\pi$ decays, assuming only that the phase
shift in the vector channel takes the value $\pi$ at threshold.
Conversely, using a small number of subtractions, the form factor may
have a significant dependence on the phase shift. Hence one can use
form factor information to learn about the phase shift itself. This is
our main goal here.

We have taken published final results of lattice simulations to
illustrate our procedure, with encouraging results. We note that
integrating this procedure more closely into the analysis of lattice
data could improve the quality of the results. This could be
especially interesting given the growing body of unquenched lattice
simulation data.

We will need a form for the phase shift as a function of the centre of
mass energy, $\sqrt s$. For this we use a scattering matrix based on
lowest-order chiral perturbation theory for the two-particle
irreducible amplitudes, which satisfies unitarity. We describe our
procedure for the case of elastic $K\pi$ scattering and subsequently
apply it to $B\pi$ and $D\pi, DK$ scattering.

\section{$\kl3$ Decays and Elastic $K\pi$ Scattering}
\label{sec:kl3-elastic}

We need calculated values of the scalar form factor for $\kl3$ decays.
For this purpose we use two-flavour dynamical domain-wall fermion
lattice results from~\cite{Dawson:2006qc}. This reference does not
contain explicit values for the chirally-extrapolated form factors
except at $q^2=0$, since the authors of~\cite{Dawson:2006qc} were
focused on calculating $f_+(0)$ accurately in order to improve the
determination of $|V_{us}|$. However it provides results for a range
of quark masses which we have used to perform our own simple chiral
extrapolation of $f_0$. Our procedure ignores correlations (details
are provided below) but we believe it leads to form factor inputs
realistic enough for use in our exploratory study. One of our
conclusions will be that accurate values of the chirally-extrapolated
form factor at a wide range of $q^2$ values are very worthwhile.

\subsection{Scalar Form Factor in $\kl3$ Decays}

The simulations in~\cite{Dawson:2006qc} were performed at a lattice
spacing $0.12\,\mathrm{fm}$ with sea quark masses in the range
$m_s/2$ to $m_s$, where $m_s$ is the physical strange quark mass.
First we used the fitted meson masses given in Table~I
of~\cite{Dawson:2006qc} to extract the physical light (up, down) and
strange quark masses. For each quark mass combination and momentum
channel ($\vec p_i \to\vec p_f$) we use the pole fit parameters given
in Table~III of~\cite{Dawson:2006qc} to determine $f_0(q^2(\vec p_i,
\vec p_f,m_{ud},m_s),m_{ud},m_s)$. Subsequently, momentum-channel by
momentum-channel, we do a linear fit in the quark masses and use this
to extract the form factor at the physical $ud$ and $s$ masses and
corresponding $q^2$ (computed using the continuum dispersion relation
and physical neutral kaon and charged pion masses). We propagate
errors by Monte Carlo through all the steps of the procedure assuming
uncorrelated Gaussian-distributed inputs (\cite{Dawson:2006qc} does
not provide correlation information). The resulting form factor values
are given in Table~\ref{tab:f0inputs}.
\begin{table}
\begin{center}
\begin{tabular}{>{$}r<{$}>{$}c<{$}}
q^2/\gev^2 & f_0(q^2)\\
\hline
0.128 & 1.0013(9)\\
0.040 & 0.957(31)\\
-0.023 & 0.914(45)\\
-0.390 & 0.883(36)\\
-0.652 & 0.817(51)
\end{tabular}
\end{center}
\caption{Input pairs $(q^2,f_0(q^2))$ obtained by the chiral
  extrapolation described in the text.}
\label{tab:f0inputs}
\end{table}
The main result of~\cite{Dawson:2006qc} is $f_+(0)=f_0(0)=0.968(9)(6)$
obtained by combining $f_+$ and $f_0$ information and using a more
realistic chiral extrapolation. We will not use this point in our fits
below since it depends on the same information as the points we do fit
and also has different systematics. We will see below that our
estimate for $f_0(0)$ is nevertheless compatible within errors,
although this is not a target of our analysis.

\subsection{Elastic $K\pi$ Scattering}

Here we show how results for $f_0(q^2)$ in Table~\ref{tab:f0inputs}
can be used to extract information on the phase shift in elastic
$s$-wave $K\pi$ scattering.

We use a multiply-subtracted Omn\`es dispersion relation to express
$f_0(q^2)$ for $q^2<\sth\equiv(m_K+m_\pi)^2$ as~\cite{NRCQM-bpi}
\begin{eqnarray}
  f_0(q^2) &=& \bigg(\prod_{i=0}^n\left[f_0(s_i)\right]^{\alpha_i(q^2)}\bigg)
  \exp\bigg\{I_\delta(q^2;\,s_0,\ldots,s_n)
             \prod_{j=0}^n(q^2-s_j)
      \bigg\}, \label{eq:omnes} \\ 
I_\delta(q^2;\, s_0,\ldots,s_n) &=&
  \frac1{\pi}\int_{\sth}^{+\infty}
  \frac{\d s}{(s-s_0)\cdots(s-s_n)}\,\frac{\delta(s)}{s-q^2},
\label{eq:phase-integral}\\
\alpha_i(s) &\equiv& \prod_{j=0, j\neq i}^n
        \frac{s-s_j}{s_i-s_j},\qquad
\alpha_i(s_j)=\delta_{ij},\qquad
\sum_{i=0}^n \alpha_i(s) = 1.
\end{eqnarray}
This representation requires as input the elastic $K\pi\to K\pi$ phase
shift $\delta(s)$ in the isospin-$1/2$ scalar channel, plus the form
factor values $\{f_0(s_i)\}$ at $n+1$ positions $\{s_i\}$ below the
$K\pi$ threshold.

The phase shift is obtained from the $K\pi$ scattering amplitude, $T$,
using
\begin{equation}
T(s) = \frac{8\pi\mi s}{\lambda^{1/2}(s,m_K^2,m_\pi^2)}
 (\e^{2\mi\delta(s)}-1)
\label{eq:Tdelta}
\end{equation}
where $s$ is the squared centre-of-mass energy and
$\lambda(x,y,z)=x^2+y^2+z^2-2(xy+yz+zx)$ is the usual kinematic
function. The (inverse) scattering amplitude, in the appropriate
isospin and angular momentum channel, is found from~\cite{EJ99,EJ99_2}
\begin{equation}
\label{eq:Tinverse}
T^{-1}(s) = -\ibar(s)-\frac1{8\pi a\sqrt\sth}
          +\frac1{V(s)}-\frac1{V(\sth)}
\end{equation}
Here, $V$ is the two-particle irreducible scattering amplitude, $a$
is the scattering length and $\ibar$ is calculated from a one-loop
bubble diagram\footnote{In the notation of reference~\cite{I&Z},
$\ibar(s) = T_G((m+M)^2)-T_G(s)$, where $M$ and $m$ are the masses of
the two propagating particles}. This description automatically
implements elastic unitarity, which is necessary for the phase shift
to be extracted from equation~(\ref{eq:Tdelta}).

For the isospin-$1/2$ scalar $K\pi$ channel, we approximate $V$ using
lowest order chiral perturbation theory (ChPT)
(see~\cite{GomezNicola:2001as} for a compilation of tree-level and
one-loop meson-meson amplitudes in ChPT):
\begin{equation}
V(s) \approx \frac1{4f_\pi^2}
  \left( m_K^2+m_\pi^2 -\frac52 s + \frac3{2s}(m_K^2-m_\pi^2)^2
  \right),
\label{eq:scattamp}
\end{equation}
where $f_\pi=92.4\mev$.

At least one subtraction is needed to make the phase-shift integral in
equation~(\ref{eq:phase-integral}) convergent. The smaller the number
of subtractions, the larger the range of $s$ over which knowledge of
the phase shift is required. There is a balance to be achieved between
the number of subtractions and the phase-shift information. Given a
fixed number of subtractions, accurate values of the form factor at
points covering a region of $q^2$ increases the knowledge of the phase
shift that can be extracted.

We start by considering a single subtraction at $q^2=0$. Hence we fit
$f_0(0)$ and the scattering length $a$ to the input form factor values
compiled in Table~\ref{tab:f0inputs}. From the chi-squared fit we
obtain
\begin{equation}
\label{eq:2paramfit}
f_0(0)= 0.950(10), \qquad
m_\pi a = 0.175(17).
\end{equation}
with $\chi^2/\mathrm{dof}=0.3$ and a correlation coefficient $-0.997$
between the two fitted parameters. The scattering length agrees well
with the experimental result $0.13$--$0.24$~\cite{Dumbrajs:1983jd} and
the one-loop $O(p^4)$ chiral perturbation theory result
$0.17(2)$~\cite{Bernard:1990kx} (both quoted in pion units). The form
factor and phase shift are displayed in Figure~\ref{fig:kl3-1sub}. The
calculated phase shift agrees remarkably well with the experimental
data up to $\sqrt s =1.2\gev$. We remind the reader that this phase
shift is obtained from the form factor via the Omn\`es relation: there
is no fit to the phase shift data itself. Despite this apparent
success, we should rule out this result. This is because we have fixed
the upper limit in the Omn\`es phase integral of
equation~(\ref{eq:phase-integral}) at $s=(1.425\gev)^2 = 5\sth$, where
typically the integrand is around three times smaller than its maximum
so that the unevaluated part of the integral may be sizeable. We have
chosen this cutoff for the integration range because the experimental
data shows the existence of additional structure above $s=(1.4\gev)^2$
(see orange points~\cite{Aston:1987ir} in lower plot of
Figure~\ref{fig:kl3-1sub}), and furthermore, $K\eta$ coupled-channel
dynamics may play a role at higher energies. These extra dynamical
features are not captured by lowest order ChPT amplitude used here.

Two approaches to address this problem are: (i) use a more realistic
model for the $K\pi$ scattering amplitude within a coupled-channel
formalism, and use higher order ChPT to determine the two-particle
irreducible amplitudes; (ii) use more subtractions to reduce the
dependence on the phase shift at large $s$ in the phase integral. The
first approach may allow the determination of low-energy constants
(LECs) appearing in higher orders in the chiral lagrangian. Although
this is an attractive prospect, given the number and accuracy of the
data points we have in the current analysis, we have not applied
higher order ChPT. However, to estimate the possible effects on the
determination of the scattering length, we have supplemented the
leading order ChPT two-particle irreducible amplitude of
equation~(\ref{eq:scattamp}) with an expression incorporating the
exchange of vector and scalar resonances as given
in~\cite{Jamin:2000wn}. This addition partially accounts for
next-to-leading ChPT contributions. We have also checked that $K\eta$
coupled-channel dynamics does not significantly affect our results. We
will note below the numerical effects on our analysis from both
massive resonances and the $K\eta$ coupled channel.

Thus we have considered two subtractions at $q^2=0$ and
$q^2=-0.75\gev^2\equiv q_1^2$ and fitted $f_0(0)$ , $f_0(q_1^2)$
and the scattering length $a$. The chi-squared fit results are:
\begin{figure}
\begin{center}
\includegraphics[width=0.7\hsize]{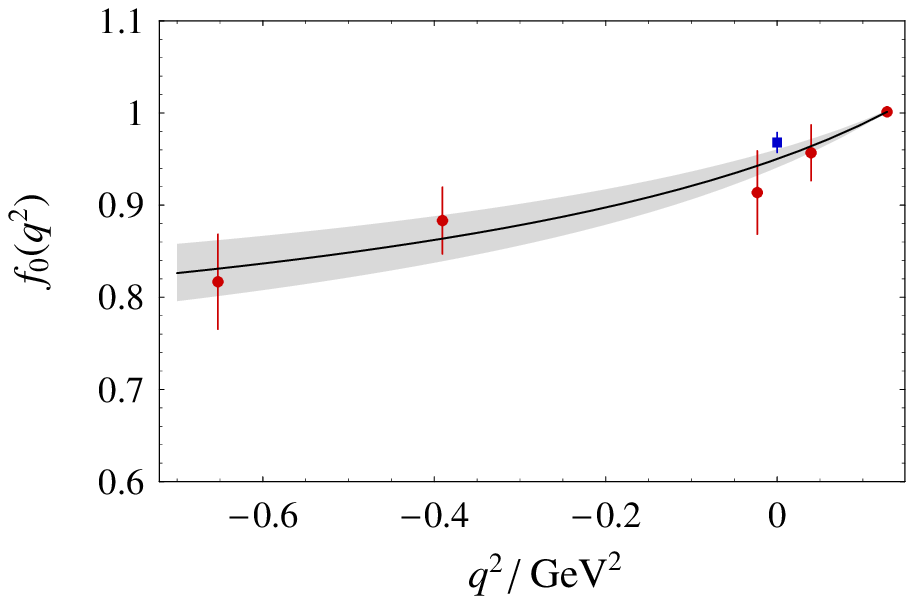}\\[1em]
\includegraphics[width=0.7\hsize]{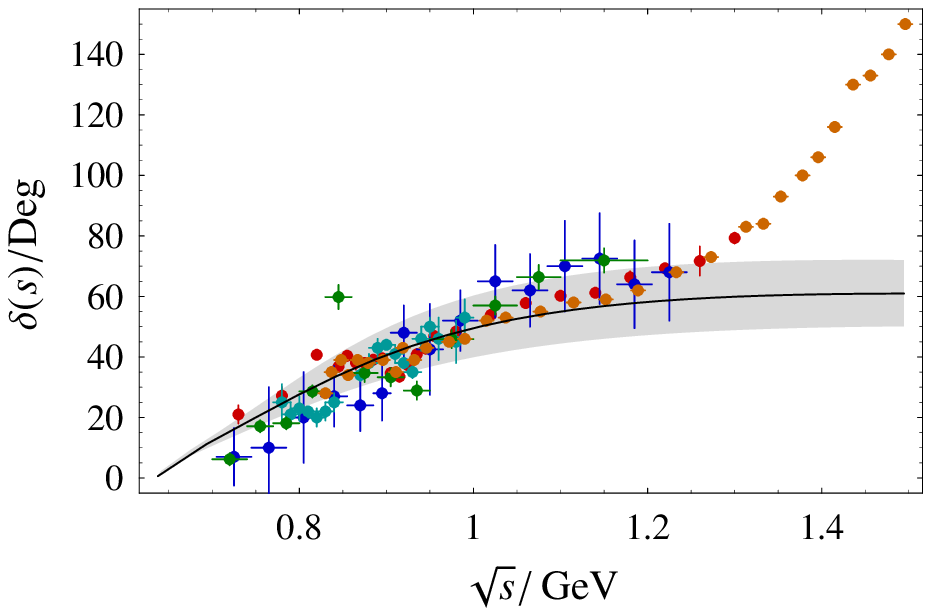}
\end{center}
\caption{The upper plot shows the $\kl3$ form factor $f_0(q^2)$, with
  a $68\%$ error band, obtained from a fit using a once-subtracted
  Omn\`es relation. Red points are the inputs from
  Table~\ref{tab:f0inputs} and the blue square shows the result
  from~\cite{Dawson:2006qc} for $f_0(0)$ (not fitted). The lower plot
  shows the isospin-$1/2$ $K\pi$ $s$-wave phase shift with a $68\%$ error
  band (grey). The phase shift plot also shows experimental data points
  from~\cite{Mercer:1971kn} (blue),%(open squares)
  \cite{Estabrooks:1977xe} (red),%(crosses)
  \cite{Bingham:1972vy} (cyan),%(solid squares)
  \cite{Baker:1974kr} (green) %(open triangles)
  and~\cite{Aston:1987ir} (orange).%(open circles)
 }
\label{fig:kl3-1sub}
\end{figure}
\begin{equation}
\label{eq:3param-fit}
\begin{array}{rcl}
f_0(0) &=& 0.946(24)\\
f_0(q_1^2) &=& 0.832(59)\\
m_\pi a &=& 0.187(95)
\end{array}
\qquad
\left(
\begin{array}{ccc}
1 & -0.517 & -0.959 \\
  & 1 & 0.738 \\
  &       & 1  \\
\end{array}
\right)
\end{equation}
with $\chi^2/\mathrm{dof}=0.4$. The fitted form factor and resulting
phase shift are illustrated in Figure~\ref{fig:kl3-2sub}.
\begin{figure}
\begin{center}
\includegraphics[width=0.7\hsize]{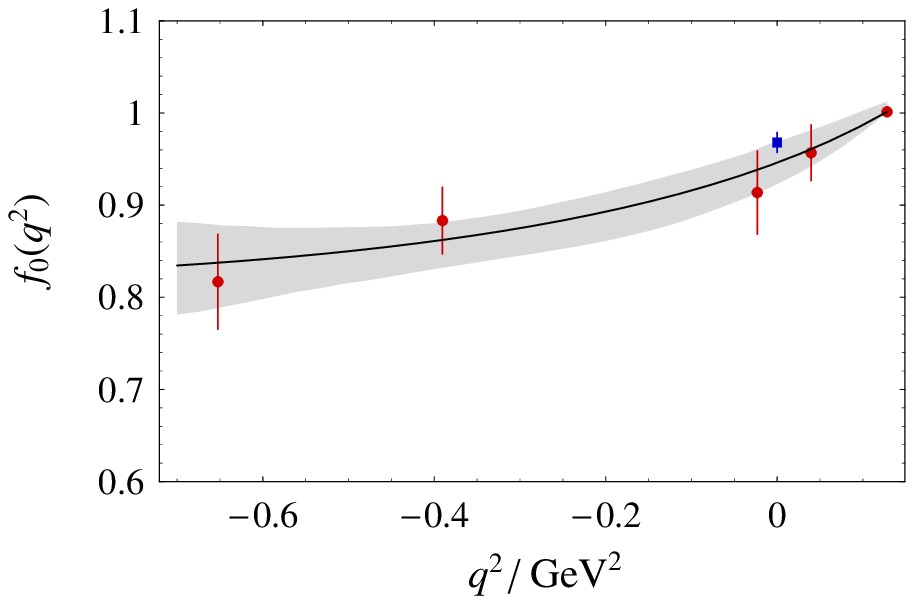}\\[1em]
\includegraphics[width=0.7\hsize]{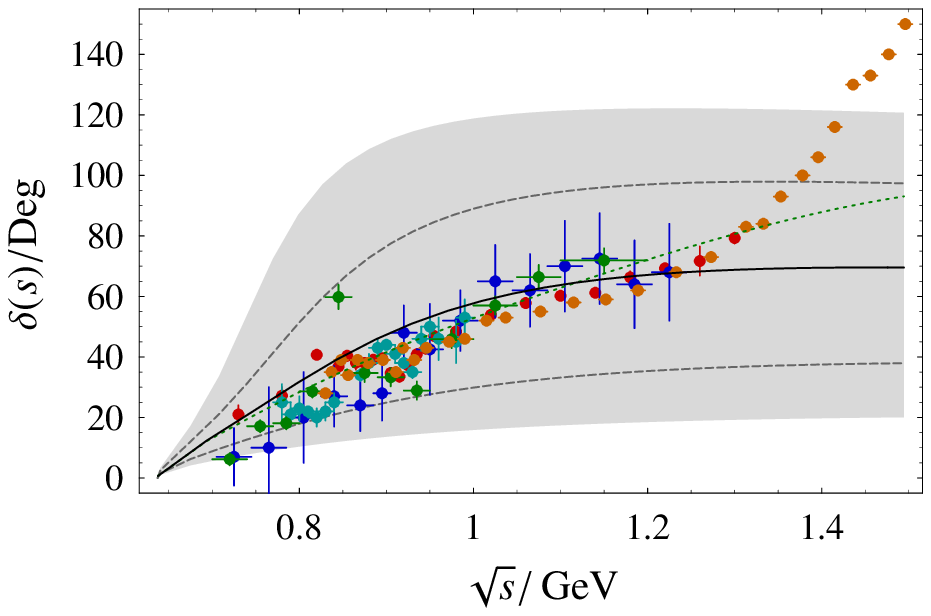}
\end{center}
\caption{The upper plot shows the $\kl3$ form factor $f_0(q^2)$, with
  a $68\%$ error band, obtained from a fit using a twice-subtracted
  Omn\`es relation. Red points are the inputs from
  Table~\ref{tab:f0inputs} and the blue square shows the result
  from~\cite{Dawson:2006qc} for $f_0(0)$ (not fitted). The lower plot
  shows the isospin-$1/2$ $K\pi$ $s$-wave phase shift with a $68\%$
  error band (grey). The green dotted line shows the phase shift
  obtained with the inclusion of resonances (see text in paragraph
  preceding discussion of the $K\eta$ coupled channel). The dashed
  lines indicate the corresponding band obtained by artificially
  halving the errror on the last four input form-factor points in
  Table~\ref{tab:f0inputs}. The phase shift plot also shows
  experimental data points as in Figure~\ref{fig:kl3-1sub}.}
\label{fig:kl3-2sub}
\end{figure}

First we note that by performing two subtractions the value of the
integrand in equation~(\ref{eq:phase-integral}) at $s=(1.4\gev)^2$, is
now typically around six times smaller than its maximum, reducing the
dependence on the phase shift at large $s$. The results presented
above were obtained with a cutoff of $9\sth=(1.9\gev)^2$ in the phase
integral, where typically the integrand is around $1/30$ of its
maximum. We have checked that raising the cutoff to infinity leads to
negligible changes in $f_0(0)$, less than $4$ parts per thousand in
$f_0(q_1^2)$ and around $2\%$ in the scattering length. These
variations are tiny compared to the statistical errors.

After the inclusion of a second subtraction, the central values of the
fitted parameters and phase shift hardly change. However, the
statistical errors on the fitted parameters have increased
significantly and in consequence the derived phase shift has larger
statistical uncertainty, as seen by comparing
Figures~\ref{fig:kl3-1sub} and~\ref{fig:kl3-2sub}. Nonetheless, this
is one of the first determinations from a lattice calculation of the
$K\pi$ scattering length, with an error comparable to the experimental
one, and of the elastic phase shift. Future more accurate lattice
simulations will improve the situation. For example, artificially
halving the errror on the last four input form-factor points in
Table~\ref{tab:f0inputs} leads to the error band indicated by the
dashed lines for the phase shift in Figure~\ref{fig:kl3-2sub} and
gives a scattering length $m_\pi a =0.187(48)$. This illustrates the
balance between the statistical uncertainty of the outputs, the number
of subtractions, knowledge of the $s$-dependence of the phase shift
and the quality of the input form factor data.

Explicit massive resonance exchanges have been ignored in our model
for the phase shift. Since the Omn\`es integration reaches $s$ values
where these contributions could be relevant, we try to estimate the
associated uncertainties. Incorporating these exchanges will also
allow us to estimate some next-to-leading ChPT
effects~\cite{Ecker:1988te,Ecker:1989yg} on the scattering length. We
obtain the resonance contribution to the isospin-$1/2$ $K\pi$
amplitude from equation~(2.4) of reference~\cite{Jamin:2000wn},
incorporating the exchange of the vector $\rho$ and $K^*$ resonances
and also of the nonet scalar mesons with masses above $1\gev$. The
vector coupling constant is fixed in the large-$N_c$
limit~\cite{Ecker:1989yg}, in good agreement with the lowest order
results coming from the decay widths of the $\rho$ and $K^*$
mesons~\cite{Ecker:1988te}. The scalar coupling constants and (common)
masses are fixed to the values in equation~(4.5)
of~\cite{Jamin:2000wn}, which lead to a reasonable description of the
isospin-$1/2$ elastic $K\pi$ scattering amplitude up to around
$1.4\gev$. When we supplement the leading order ChPT two-particle
irreducible amplitude of equation~(\ref{eq:scattamp}) with this
resonance contribution, we find no appreciable changes in the fitted
values for $f_0$ and $f_1$, while the scattering length increases by
$6\%$. For the phase shift itself, the change is much smaller than the
errors up to $1.3\gev$ with some increase in the central line to match
the data (see green dotted line in Figure~\ref{fig:kl3-2sub}). We will
account for these effects in our systematic error for the scattering
length in equation~(\ref{eq:kpi-scattlength}) below.

We have also examined $K\eta$ coupled channel effects. To do
this we have modified equation~(\ref{eq:Tinverse}) to become a
$2{\times}2$ matrix equation in the coupled channel space,
\begin{equation}
\label{eq:Tinverse-mod}
T^{-1}(s) = -\ibar(s) -C + V^{-1}(s),
\end{equation}
where the one-loop bubble integral and the coefficient matrix $C$ are
diagonal and we have approximated the $V$ matrix using lowest order
ChPT expressions for the $K\pi\to K\pi$, $K\pi\to K\eta$ and $K\eta\to
K\eta$ amplitudes\footnote{We use the Gell-Mann--Okubo relation to
express the eta mass in terms of the pion and kaon masses.} (see for
instance~\cite{GomezNicola:2001as}). In principle one should fit the
two diagonal entries in the matrix $C$, but given our limited data and
since we do not expect sizeable effects below $\sqrt s = 1.4\gev$, we
have fixed the entry corresponding to the $K\eta$ channel by demanding
that
$\big[T^{-1}(\mu)\big]_{K\eta,K\eta}=\big[V^{-1}(\mu)\big]_{K\eta,K\eta}$
for some scale
$(m_K-m_\eta)^2<\mu<(m_K+m_\eta)^2$~\cite{Lutz:2001yb,Garcia-Recio:2003ks}.
The first entry in $C$ is related to the $K\pi$ elastic scattering
length, allowing us to fit $f_0(0)$, $f_0(q_1^2)$ and $a$ as before.
There are no appreciable changes in our fitted values for $f_0(0)$ and
$f_0(q_1^2)$, while the scattering length is reduced by up to $5\%$.
In the phase shift, one can see a cusp at the $K\eta$ threshold, but
the change itself is small compared not only to the error bands in
Figure~\ref{fig:kl3-2sub} but also to those in
Figure~\ref{fig:kl3-1sub}.

In our fits we observe almost complete anticorrelation of the fitted
$f_0(0)$ and the scattering length, which means that one linear
combination of these two parameters is redundant given the current
accuracy of the input form factor information. This is not unexpected
because the lowest order ChPT expressions for $f_0(0)$ and the
scattering length are linearly related, depending only on $1/f_\pi^2$
(apart from masses). Of course, higher order corrections will not
necessarily preserve this property: according to the Ademollo--Gatto
theorem $f_0(0)$ does not have analytic terms from additional LECs in
the $O(p^4)$ chiral lagrangian~\cite{Ademollo:1964sr,Gasser:1984gg},
while such terms will affect the scattering
length~\cite{Gasser:1984gg}.

We have redone our two-subtraction fit implementing a linear relation
between $f_0(0)$ and $m_\pi a$ deduced from the one-subtraction fit
results of equation~(\ref{eq:2paramfit}), assuming that the
correlation coefficient is exactly $-1$ instead of $-0.997$. The new
results are
\begin{equation}
f_0(q_1^2) = 0.827(32),\qquad
m_\pi a = 0.179(17)
\end{equation}
with $f_0(0)=0.948(10)$ and $\chi^2/\mathrm{dof}=0.3$, with
correlation coefficient $-0.991$. Once again this strong correlation
could be used to reduce the number of fit parameters. However,
$f_0(q_1^2)$ is less-reliably calculated in ChPT and therefore its
relation to $m_\pi a$ less well-determined. Hence we proceed with
these fit results.

\begin{figure}
\begin{center}
\includegraphics[width=0.7\hsize]{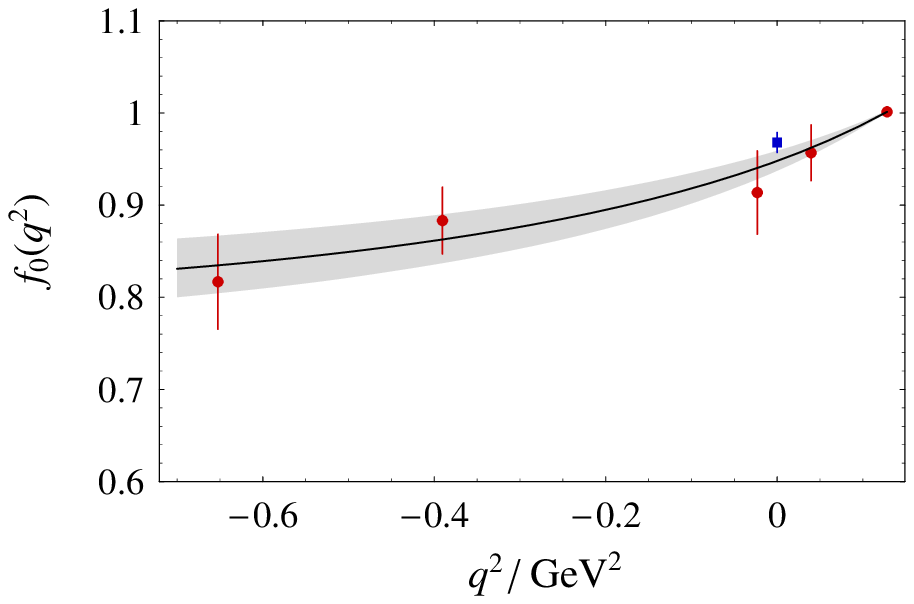}\\[1em]
\includegraphics[width=0.7\hsize]{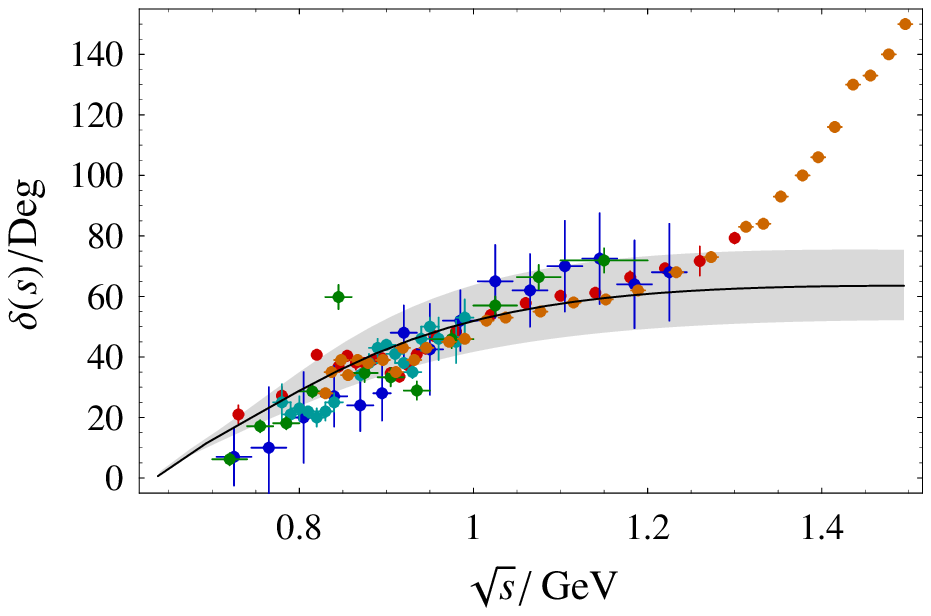}
\end{center}
\caption{The upper plot shows the $\kl3$ form factor $f_0(q^2)$, with
  a $68\%$ error band, obtained from a fit using a twice-subtracted
  Omn\`es relation, implementing a linear relation between $f_0(0)$
  and the scattering length as described in the text. Red points are
  the inputs from Table~\ref{tab:f0inputs} and the blue square shows
  the result from~\cite{Dawson:2006qc} for $f_0(0)$ (not fitted). The
  lower plot shows the isospin-$1/2$ $K\pi$ $s$-wave phase shift with
  a $68\%$ error band (grey). The phase shift plot also shows
  experimental data points as in Figure~\ref{fig:kl3-1sub}.}
\label{fig:kl3-2sub-corr}
\end{figure}

The fitted form factor and derived phase shift are shown in
Figure~\ref{fig:kl3-2sub-corr}. Both the scattering length and derived
phase shift below $\sqrt s=1.4\gev$ are in remarkable agreement with
experiment, with a $10\%$ error for the scattering length and
$10$--$20\%$ for the phase shift. The scattering length also agrees
with the recent lattice finite-volume effect result
$0.1725^{+0.0029}_{-0.0157}$~\cite{Beane:2006gj} and with the one-loop
ChPT result $0.17(2)$~\cite{Bernard:1990kx}. Following the discussion
above on massive resonance exchanges and $K\eta$ coupled channel
effects, we conclude that they are covered by the statistical error
bands in the phase shift, but we combine their effects in quadrature
and include an $8\%$ systematic error in our final result for the
scattering length:
\begin{equation}
\label{eq:kpi-scattlength}
m_\pi a = 0.179(17)(14).
\end{equation}

We expect that more precise, accurate form factor data will be more
sensitive to higher-order ChPT corrections, making the $\chi^2$
fully-dependent on both $f_0(0)$ and $m_\pi a$ and moving their
correlation away from $-1$. In that situation we will not need to
implement the procedure above, since a three-parameter (two
subtractions) fit should provide smaller errors than we saw in
equation~(\ref{eq:3param-fit}).

\section{Semileptonic $H\to\pi,K$ Decays and $s$-wave Elastic Scattering
  Phase Shifts}

\subsection{$B\pi$ Scalar Form Factor and Phase Shifts}

To discuss the $B\pi$ phase shifts we take over the formalism
described above, with appropriate changes to the kinematics,
considering a neutral $B$ meson and charged pion. For the two-particle
irreducible isospin-$1/2$ $s$-wave $B\pi$ scattering amplitude we use
the leading contact term from the heavy meson chiral perturbation
theory (HMChPT) lagrangian~\cite{Wise:1992hn},
\begin{equation}
\label{eq:hmchpt-amp}
V(s) \approx \frac1{4f_\pi^2}
 \left( 2(m_B^2+m_\pi^2)-3s +\frac{(m_B^2-m_\pi^2)^2}s
 \right).
\end{equation}
We have not included a contribution from the $t$-channel
$B^*$-exchange diagram depending on the leading HMChPT $B^*B\pi$
interaction term, since this vanishes at $\sth$ and has magnitude less
than $1\%$ of that from the expression above over a large range of
$s$.

We take input scalar form factor values from the lattice QCD
calculation by the HPQCD collaboration~\cite{Gulez:2006dt}, assuming
that the statistical errors, $\sigma_i$, are uncorrelated, while the
quoted $9\%$, $3\%$ and $1\%$ systematic errors are combined to give
an additional $10\%$ fully-correlated error, $\epsilon_i$, on each
point. The input covariance matrix for the lattice data thus takes the
form $C_{ij} = \sigma_i^2 \delta_{ij} + \epsilon_i\epsilon_j$. We also
use the lightcone sumrule result for $f_0(0)=f_+(0)$
from~\cite{LCSR_04_BZ}. These inputs are collected in
Table~\ref{tab:bpi-inputs}.
\begin{table}
\begin{center}
\begin{tabular}{l>{$}c<{$}>{$}c<{$}}
%\hline
%\vrule height2.5ex depth0pt width0pt
 & q^2/\gev^2 & f_0(q^2) \\
\hline
LCSR~\cite{LCSR_04_BZ} & 0 & 0.258\pm0.031 \\
\hline
HPQCD~\cite{Gulez:2006dt}
 & 15.23 & 0.475\pm0.026 \\
 & 16.28 & 0.508\pm0.025 \\
 & 17.34 & 0.527\pm0.025 \\
 & 18.39 & 0.568\pm0.024 \\
 & 19.45 & 0.610\pm0.024 \\
 & 20.51 & 0.651\pm0.025 \\
 & 21.56 & 0.703\pm0.026 %\\
%\hline
\end{tabular}
\end{center}
\caption{$B\pi$ scalar form factor inputs. A fully-correlated $10\%$
  systematic error should be added to the statistical error listed in
  the table for the HPQCD points.}
\label{tab:bpi-inputs}
\end{table}

\begin{figure}
\begin{center}
\includegraphics[width=0.7\hsize]{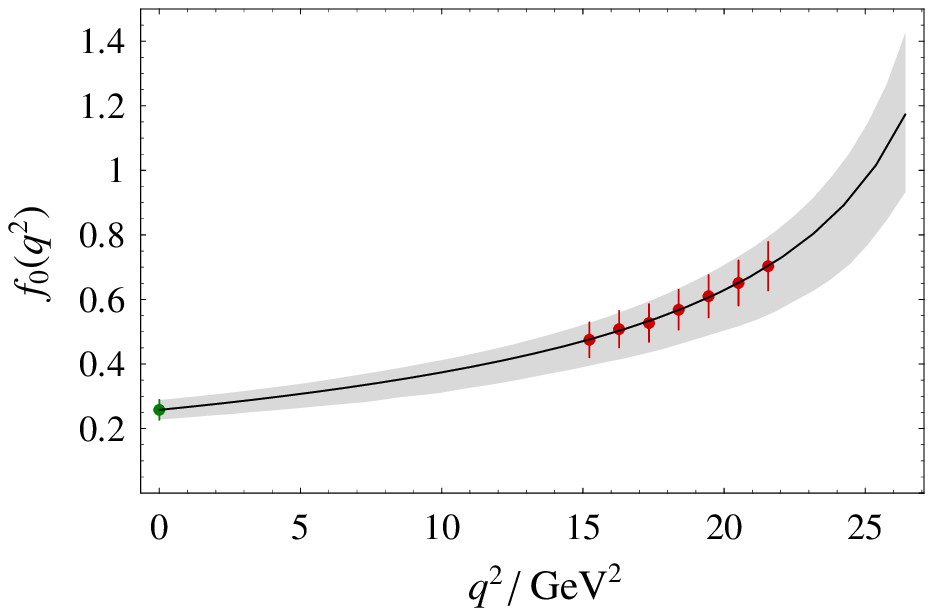}\\[1em]
\includegraphics[width=0.7\hsize]{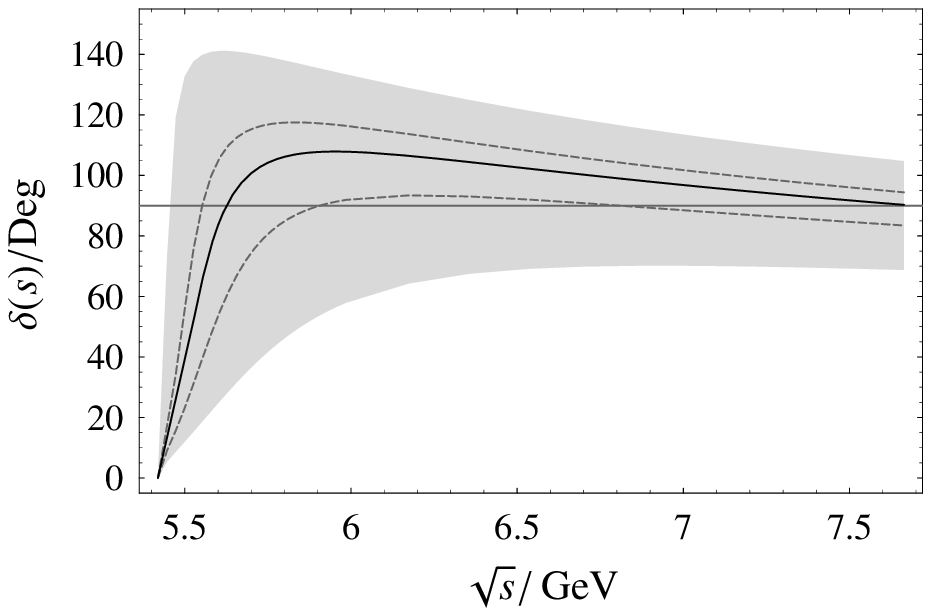}
\end{center}
\caption{$B\pi$ isospin-$1/2$ scalar form factor and phase shift,
together with $68\%$ confidence level bounds (grey bands). The
points on the form factor plot are the inputs given in
Table~\ref{tab:bpi-inputs}. The dashed curves on the phase
shift plot show the effect on the statistical uncertainty of
reducing the input errors to $1/4$ of their current value. The
intercept of the phase shift with the horizontal line at
$90^\circ$ indicates the position of a resonance.}
\label{fig:bpi}
\end{figure}

Given the large mass of the $B$ meson, the influence of inelastic
single and multiple light meson production may be important within a
few hundred $\mev$ of threshold. To reduce the impact of these
inelastic channels in the Omn\`es phase integral, we use two
subtractions at $q^2=0$ and $q^2=\qsqmax=(m_B-m_\pi)^2$. Hence we have
performed a three-parameter fit to $f_0(0)$, $f_0(\qsqmax)$ and the
scattering length. The fit results are:
\begin{equation}
\label{eq:bpi-3param-fit}
\begin{array}{rcl}
f_0(0) &=& 0.258(31)\\
f_0(\qsqmax) &=& 1.17(24)\\
m_\pi a &=& 0.26(26)
\end{array}
\qquad
\left(
\begin{array}{ccc}
1 & 0.448 & 0.532 \\
  & 1 & 0.814 \\
  &       & 1 \\
\end{array}
\right)
\end{equation}
with $\chi^2/\mathrm{dof}=0.03$. We show the fitted form factor and
the derived phase shift in Figure~\ref{fig:bpi}. We have integrated up
to $s=5\sth\approx (12.1\gev)^2$ where the integrand is typically one
thousandth of its maximum value (reached at $s=(5.6\gev)^2$ when the
integral is evaluated for $q^2=\qsqmax/2$). We plot the phase shift up
to $\sqrt s=7.5\gev$ where the integrand is already $30$ times smaller
than its maximum value. We also observe that the fitted value for
$f_0(\qsqmax)$ agrees within errors with the heavy quark effective
theory prediction in the soft-pion limit~\cite{Burdman:1993es},
$f_0(m_B^2) = f_B/f_\pi + \mathcal{O}(1/m_b^2) \approx 1.4(2)$ (using
$f_B=189(27)\mev$~\cite{Hashimoto:2004hn}).

We find that it is possible to determine the scattering length and the
phase shift from current lattice QCD and sumrule form factor
calculations, albeit with large errors. Moreover, we observe that our
central phase-shift curve shows the existence of a resonance at $\sqrt
s\approx5.6\gev$ which may have some experimental
support~\cite{Ciulli:1999ig}. However, with the current level of
errors in the form factor inputs we cannot give an upper bound for
this resonance mass. The dashed lines in the phase shift plot show the
effect of reducing the input errors to $1/4$ of their current size,
comparable to those for the $K_{l3}$ results. In this case we would be
able to constrain the resonance mass, as can be seen from the figure.
These reduced input errors would lead to a determination of the
scattering length with $25\%$ error. Having more input points, or
indeed a functional form (as we use below for the $D\pi$ and $DK$
cases), would of course also reduce the uncertainties.

\subsection{$D\pi$ and $DK$ Phase Shifts}

To discuss the $D\pi$ phase shift we will use
equation~(\ref{eq:hmchpt-amp}) with the obvious replacement $m_B\to
m_D$. For the $DK$ phase shift we project into the isospin zero
channel, where the two-particle irreducible amplitude again takes the
same form with the appropriate substitutions of masses and the
replacement $f_\pi \to f_K\approx 110\mev$.

We take input scalar form factor values from the Fermilab-MILC-HPQCD
lattice QCD calculation of reference~\cite{Aubin:2004ej}. The chiral
extrapolation procedure adopted there leads to parameters for a
Becirevic-Kaidalov (BK)~\cite{Becirevic:1999kt} parameterisation of
$f_0(q^2)$, and hence an explicit functional form is determined,
rather than values at a set of $q^2$ points. For $f_0(q^2)$, the BK
function is a simple pole form
\begin{equation}
f_0^\mathrm{BK}(q^2) = \frac F{1-q^2/(\beta m_{D_x^*}^2)}.
\end{equation}
where $m_{D_x^*}=2.010\gev,\, 2.112\gev$ for $D\pi, DK$ respectively.
The BK parameters $F$ and $\beta$ are compiled in Table~I
of~\cite{Aubin:2004ej} and repeated here:
\begin{equation}
\begin{array}{r<{,\qquad}r<{.}}
F_{D\pi}=0.64(3) & \beta_{D\pi} = 1.41(6)\\
F_{DK}  =0.73(3) & \beta_{DK}   = 1.31(7)
\end{array}
\end{equation}
The errors above are statistical. We have added in quadrature a
further $10\%$ error to the $F$ parameter to account for the
systematic uncertainty for the form factors quoted
in~\cite{Aubin:2004ej}.

The fitting procedure we use here is as follows. We assume that $F$
and $\beta$ are uncorrelated\footnote{Reference~\cite{Aubin:2004ej}
does not provide correlation information for the fitted parameters.}
Gaussian-distributed variables. We perform a Monte Carlo procedure by
generating an ensemble of $\{F,\beta\}$ pairs. For each pair, we
determine two Omn\`es subtraction parameters $f_{1} \equiv f_0(0)$,
$f_2 \equiv f_0(\qsqmax)$ and the scattering length $a$ by minimising
\begin{equation}
\label{eq:norm}
\int_{-0.5\gev^2}^{q_\mathrm{upper}^2} \d q^2
 \left| f^{\mbox{\scriptsize Omn\`es}}_0(q^2;a,f_1,f_2) -
        f^\mathrm{BK}_0(q^2;F,\beta)
 \right|^2
\end{equation}
where $f_0^{\mbox{\scriptsize Omn\`es}}$ is easily obtained from
equation~(\ref{eq:omnes}). For $D\pi$ we take
$q_\mathrm{upper}^2=2\gev^2$, while for $DK$,
$q_\mathrm{upper}^2=(m_D-m_K)^2$ (these choices correspond to the
$q^2$ ranges shown in Figure~(3) of~\cite{Aubin:2004ej}). This
produces a three dimensional distribution for $f_1$, $f_2$ and $a$.

For $D\pi$ we determine the following central values and correlation
matrix:
\begin{equation}
\label{eq:dpi-3param-fit}
\begin{array}{rcl}
f_0(0) &=& 0.64(7)\\
f_0(\qsqmax) &=& 1.38(17)\\
m_\pi a &=& 0.29(4)
\end{array}
\qquad
\left(
\begin{array}{ccc}
1 & 0.9 & 0.0 \\
  & 1   & 0.4 \\
  &     & 1  \\
\end{array}
\right)
\end{equation}
Since we are fitting a three-parameter function to input form-factors
determined by two parameters, we find that the correlation matrix has
determinant compatible with zero. This feature could be avoided by
using the Omn\`es parameterisation throughout the analysis.

\begin{figure}
\begin{center}
\includegraphics[width=0.7\hsize]{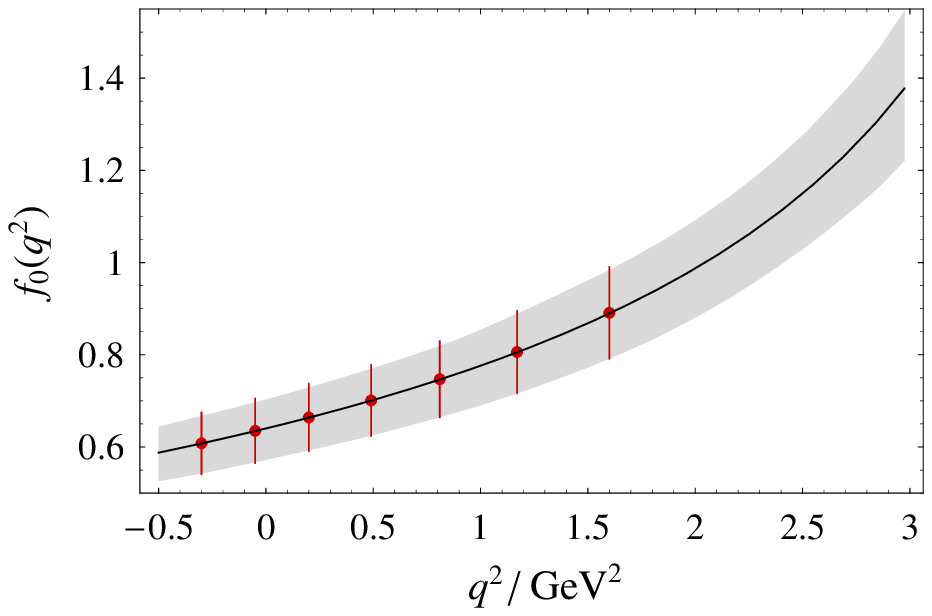}\\[1em]
\includegraphics[width=0.7\hsize]{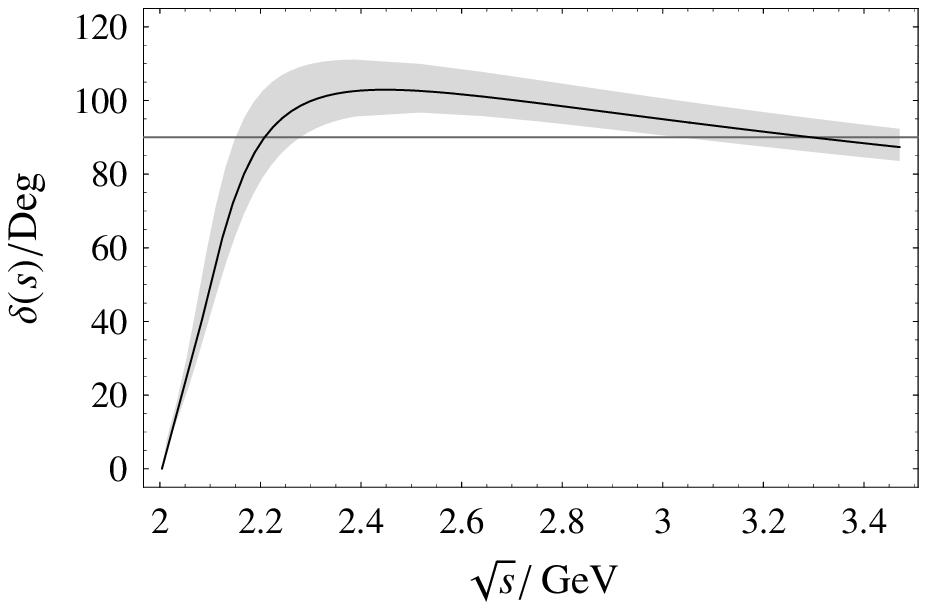}
\end{center}
\caption{$D\pi$ isospin-$1/2$ scalar form factor and phase shift,
together with $68\%$ confidence level bounds. The points on the form
factor plot are read off from Figure~3 of~\cite{Aubin:2004ej}, but
with error bars expanded to include a $10\%$ systematic error, and are
displayed to show the good agreement with our fit. The intercept of
the phase shift with the horizontal line at $90^\circ$
indicates the position of a resonance.}
\label{fig:dpi-fit}
\end{figure}

We show our fitted form factor and derived phase shift in
Figure~\ref{fig:dpi-fit}. We have integrated the phase shift in the
Omn\`es integral up to $s=5\sth\approx (4.5\gev)^2$ where the
integrand is typically one $250$th of its maximum value (reached at
$s=(2.2\gev)^2$ when the integral is evaluated for $q^2=\qsqmax/2$).
We plot the phase shift up to $\sqrt s=3.5\gev$ where the integrand is
already $40$ times smaller than its maximum value.

The fitted form factor is indistinguishable by eye from the input BK
curve. Since the Omn\`es expression for the form factor is founded
only on very general properties, we observe that the BK
parameterisation could be replaced by the Omn\`es form throughout an
analysis of lattice data, removing the need for a fit like the one
done above. We remark that the scattering length can be determined
with a small error and differs from the value $m_\pi a =0.18$ found
from lowest order HMChPT. We believe that the small errors result from
fitting a functional form rather than a small set of points. We
predict the existence of an $I=1/2$ $s$-wave resonance at
$2.2(1)\gev$.

When we consider the $DK$ channel we find that for almost all of our
Monte Carlo trials, the fitted value for the scattering length is
huge, effectively infinite. This tells us that
$\mathrm{Re}T^{-1}(\sth)=0$ as can be seen from
equation~(\ref{eq:Tinverse}). Hence there should be a resonance at
threshold, $(m_D+m_K)^2 = (2.36\gev)^2$. This can be understood by
noting the existence of a $0^+$ state, $D_{s0}^+(2317)$, discovered by
Babar~\cite{Aubert:2003fg}, which is likely an isoscalar~\cite{pdg}.
Neglecting isospin-violating decays to $D_s^+\pi^0$, this state could
be considered as an isoscalar $s$-wave $DK$ bound state. In this case,
following Levinson's theorem~\cite{MS70}, the phase shift close to
threshold has the form $\pi + p a +\cdots$, where $p$ is the
centre-of-mass three-momentum. Using three parameter fits (two
subtractions and $a$) we find that the scattering length is
effectively zero in $70\%$ of our Monte Carlo trials. Given this, we
assume that the phase shift is $\pi$ over the range where the
integrand of the phase-shift integral is significant and obtain an
excellent two-parameter fit (two subtractions):
\begin{equation}
\label{eq:dk-2param-fit}
\begin{array}{rcl}
f_0(0) &=& 0.73(9)\\
f_0(\qsqmax) &=& 1.08(12)\\
\end{array}
\end{equation}
with correlation coefficient $0.975$.

\section{Conclusions}

We have shown how existing theoretical, principally lattice,
calculations of the scalar form factors in semileptonic
pseudoscalar-to-pseudoscalar decays can be used to extract information
about the phase shifts in the corresponding elastic $s$-wave
scattering channels. The Omn\`es expression for the form factor rests
on general principles of analyticity and unitarity. We remark that it
provides a model-independent functional form that can be used in
analysing lattice data, replacing more phenomenological
parameterisations. Using the Omn\`es expression throughout would allow
correlations in the lattice data to be taken into account.

From $K_{l3}$ decays we have determined the elastic $K\pi$ scattering
length with an uncertainty of around $12\%$ and find a reasonable
description of phase-shift data up to $1.3\gev$. Improved form-factor
data could potentially be used to learn information about low-energy
constants (LECs) in the $O(p^4)$ chiral lagrangian.

For $B\pi$ the extracted scattering length and phase shift suffer from
large uncertainties. We found hints of a resonance around $5.6\gev$.
Reduced errors on the input form factor points, or increasing the
number of points, could enable the position of the resonance to be
established.

For $D\pi$ and $DK$ we took advantage of functional forms for the
$f_0(q^2)$ arising from lattice simulations. For $D\pi$ we were able
to extract the scattering length with $13\%$ statistical error and
found a phase shift showing the existence of a resonance at around
$2.2\gev$. The scattering length so determined is around $70\%$ higher
than that predicted by lowest order HMChPT. For $DK$ we found hints
that there is a bound state which could be identified with the
$D_{s0}^+(2317)$.

\subsubsection*{Acknowledgements}

We thank Feng-Kun Guo and JA~Oller for helpful correspondence. JMF
acknowledges the hospitality of the Departamento de F\'isica
At\'omica, Molecular y Nuclear, Universidad de Granada, MEC support
for estancias de Profesores e investigadores extranjeros en r\'egimen
de a\~no sab\'atico en Espa\~na SAB2005--0163, and PPARC grant
PP/D000211/1. JN acknowledges the hospitality of the School of Physics
\& Astronomy at the University of Southampton, Junta de Andalucia
grant FQM0225, MEC grant FIS2005--00810 and MEC financial support for
movilidad de Profesores de Universidad espa\~noles PR2006--0403. JMF
and JN acknowledge support from the EU Human Resources and Mobility
Activity, FLAVIAnet, contract number MRTN--CT--2006--035482.

\bibliographystyle{physrev}
\bibliography{omnes2}

\end{document}